# On photonic tunnelling and the possibility of superluminal transport of electromagnetic energy

Luca Nanni

luca.nanni@edu.unife.it

**Abstract**. Motivated by increased interest in experiments in which light appears to propagate by tunnelling at superluminal velocity, the Lorentz invariant theory proposed by Partha Ghose to explain these surprising effects is revisited. This theory is based on the Harish-Chandra formalism, which describes the relativistic dynamics of a massless spin-1 boson, like a photon. Via this formalism, the Bohmian average transport velocity of the electromagnetic energy is formulated. It is proved that, if the dielectric making the waveguide is nonabsorptive and nondispersive, this velocity can be superluminal. This result is validated in the framework of quantum electrodynamics, demonstrating that the average velocity of the photon inside the waveguide is given by the contribution of instantaneous superluminal velocities. This theory, therefore, suggests the optimal conditions for designing optical devices capable of locally transporting electromagnetic energy at superluminal velocities mitigating the signal attenuation.

## 1. Introduction

In recent decades, several impressive experiments have been performed in the framework of quantum optics, claiming that electromagnetic waves may be propagated faster than light [1-9]. Superluminal behaviour is particularly manifested when such experiments are carried out using dispersive media, waveguides, and devices based on the tunnelling effect. This has opened a controversial debate about the physical meaning and interpretation of the results, which seem to violate the principle of causality [10-15]. In this scenario, theoretical physicists have worked intensely to find a quantum theory capable of correctly interpreting the experimental results, corroborating or refuting the experimental physicists' conclusions [15-21]. Without moving into speculative fields, the Lorentz invariant theories seem to be worthwhile foundations for interpreting the experimental results and preserving the principle of causality [22-25]. A source of confusion in interpreting these results stems from the inappropriate use of group velocity, phase velocity, and signal front velocity, which have been clearly differentiated since the early 1900s and have led to the conclusion that no signal can propagate at velocities higher than $c$ [26-27]. However, the superluminality of physical signals of finite duration in dispersive media can be explained, without violating the principle of causality, via Sudarshan's reinterpretation principle [23]. Nevertheless, in cases of superluminal photon tunnelling, Sudarshan's theory fails because, within the photonic barrier, the signal becomes an exponentially damped evanescent wave [28]. Evanescent waves are localised in space and do not accumulate any phase while propagating within the optical medium. In such a case, the problem is determining how to explain the propagation of the photon inside the barrier with zero phase delay. This problem has often been addressed by the Schrodinger equation [18-19,29-30], which, in cases of one-dimensional non-relativistic tunnelling, is reduced to the Helmholtz equation. The latter coincides with Maxwell's equations in an isotropic and homogeneous medium. The obtained model, thus, explains the superluminal dynamics of photonic tunnelling but in an unrealistic framework, since the Schrodinger equation, being non-relativistic, cannot guarantee the Lorentz invariance and the consequent causality principle. Moreover, this theoretical approach leads to a complex wave function that is incompatible with the real Helmholtz function for the electromagnetic field. The gaps lead to a complete revision of the theory, which, in our opinion, must instead be addressed in the framework of the relativistic formalism of photons.

This work starts from a brilliant idea by Partha Ghose (PG) which dates back to 2001 and could be useful for physicists who deal with superluminal phenomena in the field of photonics and plasmonics, in order to refine the models already proposed in the literature [31]. We study the tunnelling of a photon through a thin, nonabsorptive, nondispersive and non-magnetic material with a band gap around the frequency ω, according to the formalism of Harish-Chandra (H-C), which was created to describe the dynamics of massless spin-1 particles [32]. This formalism is relativistic and leads to a real, ten-component wave function, of which the first six non-trivial components separately satisfy Maxwell's equations and the last four are identically zero. It shows that, with an appropriately chosen photonic material forming the optical device, electromagnetic energy can be transported at a velocity greater than $c$, without violating the principle of causality. The average tunnelling velocity is calculated using the energy flux current, thus avoiding the misleading results obtained with the phase gradient, as occurs in the non-relativistic approach. This result is validated in the framework of quantum electrodynamics, proving that the transition probability amplitude in a whatever spacelike interval within the waveguide is always different from zero. This means that the instantaneous transport velocity of the electromagnetic energy into the media is always superluminal and, therefore, an average velocity greater than the speed of light is expected. Our theory corroborates the experiments mentioned at the beginning of this section – at least those performed using optical devices based on quantum tunnelling. A further advantage of this approach is that it never refers to phase velocity or group velocity, least of all to ambiguous definitions of tunnelling time [33]. The proposed model places constraints on the design of optical devices leading to superluminal tunnelling and could suggest useful ideas for improving the performance of new experiments to confirm the previous results.

**2. The Partha Ghose Approach to Superluminal Tunnelling**

This section the approach to superluminal photon tunnelling proposed by PG [31] is revised, in order to prepare for the treatment of the problem which will be addressed in the next section. The PG approach is based on the H-C formalism, which describes in relativistically invariant way the dynamics of a massless spin-1 bosons. The H-C equation, obtained from that of Kemmer formulated for massive spin-1 mesons [34], reads

$$(i\hbar\beta_\mu\partial^\mu + mc\Gamma)\psi = 0, \qquad (1)$$

where $\mu = 0,1,2,3$, $\psi$ is the Kemmer wave function for spin-1 meson, $\beta_\mu$ and $\Gamma$ 10x10 are matrices satisfying the algebraical relation:

$$\Gamma^2 = \Gamma \quad \text{and} \quad \Gamma\beta_\mu + \beta_\mu\Gamma = \beta_\mu. \qquad (2)$$

The explicit forms of $\beta_\mu$ matrices are

$$\begin{cases} \beta_0 = i\begin{pmatrix} 0_{3\times3} & 0_{3\times3} & \bar{\mathbb{1}}_{3\times3} & 0_{\times3} \\ 0_{3\times3} & 0_{3\times3} & 0_{3\times3} & 0_{\times3} \\ \mathbb{1}_{3\times3} & 0_{3\times3} & 0_{3\times3} & 0_{\times3} \\ 0_{3\times} & 0_{3\times} & 0_{3\times} & 0 \end{pmatrix} & \beta_1 = i\begin{pmatrix} 0_{3\times3} & 0_{3\times3} & 0_{3\times3} & \bar{e}_1 \\ 0_{3\times3} & 0_{3\times3} & A_{3\times3} & 0_{\times3} \\ \mathbb{1}_{3\times3} & \bar{A}_{3\times3} & 0_{3\times3} & 0_{\times3} \\ \bar{e}_1{}^t & 0_{3x} & 0_{3x} & 0 \end{pmatrix} \\ \beta_2 = i\begin{pmatrix} 0_{3\times3} & 0_{3\times3} & 0_{3\times3} & \bar{e}_2 \\ 0_{3\times3} & 0_{3\times3} & B_{3\times3} & 0_{\times3} \\ \mathbb{1}_{3\times3} & \bar{B}_{3\times3} & 0_{3\times3} & 0_{\times3} \\ \bar{e}_2{}^t & 0_{3\times} & 0_{3\times} & 0 \end{pmatrix} & \beta_3 = i\begin{pmatrix} 0_{3\times3} & 0_{3\times3} & 0_{3\times3} & \bar{e}_3 \\ 0_{3\times3} & 0_{3\times3} & C_{3\times3} & 0_{\times3} \\ \mathbb{1}_{3\times3} & \bar{C}_{3\times3} & 0_{3\times3} & 0_{\times3} \\ \bar{e}_3{}^t & 0_{3\times} & 0_{3\times} & 0 \end{pmatrix} \end{cases}, \qquad (3)$$

where $0_{3\times3}$ is the 3x3 zero matrix; $0_{\times3}$ is the three-component zero column vector; $0_{3\times}$ is the three-component zero row vector; $\mathbb{1}_{3\times3}$ is the 3x3 unity matrix; $\bar{\mathbb{1}}_{3\times3} = -\mathbb{1}_{3\times3}$; $\bar{e}_1 = -(1,0,0)$; $\bar{e}_2 = -(0,1,0)$; $\bar{e}_3 = -(0,0,1)$; and

$$A_{3\times3} = \begin{pmatrix} 0 & 0 & 0 \\ 0 & 0 & \bar{1} \\ 0 & 1 & 0 \end{pmatrix} \quad B_{3\times3} = \begin{pmatrix} 0 & 0 & 1 \\ 0 & 0 & 0 \\ \bar{1} & 0 & 0 \end{pmatrix} \quad C_{3\times3} = \begin{pmatrix} 0 & \bar{1} & 0 \\ 1 & 0 & 0 \\ 0 & 0 & 0 \end{pmatrix}. \qquad (4)$$

As usual, $\bar{A}_{3\times3} = -A_{3\times3}$; $\bar{B}_{3\times3} = -B_{3\times3}$; and $\bar{C}_{3\times3} = -C_{3\times3}$. As explained by Harish-Chandra, in Eq. (1) the term of mass, not consistent with the photon, is present only for reasons of unit of measure and disappears when calculating other dynamic operators associated with physical observables. The meaning of the matrix $\Gamma$ will be clarified shortly. By appropriate manipulations, detailed in the reference [34], Eq. (1) can be rewritten as

$$i\hbar\beta_i\beta_0{}^2\partial_i\psi + mc(\mathbb{1} - \beta_0{}^2)(\Gamma\psi) = 0. \qquad (5)$$

For the photon, which is its own antiparticle, the wave function $(\Gamma\psi)$ must be formed by ten real components, of which the last four are zero. Furthermore, each of the nontrivial components must separately satisfy Maxwell's equations. With $\psi$ being the wave function of the spin-1 massive meson, we can now understand how the matrix $\Gamma$ acts to transform it in a wave function with the required properties. More precisely, $\Gamma$ transforms $\psi$, making each component real and setting to zero the last four components of the meson spinor which, being massive, are not null. A good choice for the wave function $(\Gamma\psi)$ is:

$$(\Gamma\psi) = (mc^2)^{-1/2}(-\boldsymbol{E}, \boldsymbol{H}, 0,0,0,0), \tag{6}$$

where $\boldsymbol{E}$ and $\boldsymbol{H}$ are the electric and magnetic field vectors in vacuum, respectively. Inside the optical barrier, these vectors will be denoted, respectively, by $\boldsymbol{D}$ and $\boldsymbol{B}$ and are given by $\boldsymbol{D} = \varepsilon_0 \boldsymbol{E} + \boldsymbol{P}$, where $\boldsymbol{P}$ is the polarization vector, and $\boldsymbol{B} = \mu_0 \boldsymbol{H} + \boldsymbol{M}$, where $\boldsymbol{M}$ is the magnetization vector, respectively. Using Eq. (5) and Eq. (6), PG obtains the temporal component of the symmetric energy-momentum tensor

$$T_{00} = -\frac{mc^2}{2}\psi^\dagger \Gamma \psi. \tag{7}$$

where $mc^2 \psi^\dagger \Gamma \psi = (\boldsymbol{D}^2 + \boldsymbol{B}^2)$ inside the optical barrier. The other components $T_{0\nu}$ are instead given by

$$T_{01} = \frac{mc^2}{2}\psi^\dagger \Gamma \beta_1 \Gamma \psi \; ; \; T_{02} = \frac{mc^2}{2}\psi^\dagger \Gamma \beta_2 \Gamma \psi \; ; \; T_{03} = \frac{mc^2}{2}\psi^\dagger \Gamma \beta_3 \Gamma \psi \tag{8}$$

In the framework of De Broglie-Bohm quantum mechanics, the velocity of energy transport can be easily obtained [35]

$$\boldsymbol{u} = c\frac{\psi^\dagger \Gamma \boldsymbol{\beta} \Gamma \psi}{\psi^\dagger \Gamma \psi} = -c\frac{T_{01} + T_{02} + T_{03}}{T_{00}}. \tag{9}$$

where $\boldsymbol{\beta}$ is the vector $(\beta_1, \beta_2, \beta_3)$. As expected, neither $-T_{00}$ nor $\boldsymbol{u}$ contain the mass term $mc^2$, confirming what has been anticipated. The term $\psi^t \Gamma \boldsymbol{\beta} \Gamma \psi$ in Eq. (9) is related to the Poynting vector. In fact, using the explicit form of $\boldsymbol{\beta}$ and $(\Gamma\psi)$, inside the optical barrier one obtains $mc^3(\psi^t \Gamma \boldsymbol{\beta} \Gamma \psi) = c(\boldsymbol{D}x\boldsymbol{B}) = \boldsymbol{S}$, where $\boldsymbol{S}$ is the Poynting vector. It must be noted that Eq. (9) depends on the energy flux current rather than on the phase gradient. This allows photonic tunnelling to be studied without using ambiguously defined quantities, such as dwell time or phase time [10-11].

## 3. Photon Tunnelling: the Model

Let us consider the tunnelling of a photon through a thin, non-absorptive, non-dispersive and non-magnetic material with a band gap of frequency $\omega_c$. The length of the potential barrier is $L$, and the photon normally hits the optical material. Photon propagation takes place along one dimension only. From now on we will define this system as waveguide or optical barrier. Photons with frequencies lower than $\omega_c$ can be transmitted within the optical media only by quantum tunnelling. The reference frame is set at the beginning of the barrier in such a way that the propagation occurs along the x-axis. Therefore, the oscillating electric field vector outside the optical barrier – i.e., for $x < 0$ and $x > L$ – is $\boldsymbol{E} = (0,0,E_z)$. The oscillating function $E_z$ is expressed by a wave packet pulse, and its explicit form is

$$E_z = \int f(k) exp\{i(\omega t - kz)\} dk, \qquad (10)$$

where $f(k)$ is the envelop function of the wave packet, $\omega$ and $k$ the angular frequency and the wave vector respectively ($k = k_z$ since $\boldsymbol{E} = (0,0,E_z)$). For simplicity let us suppose $f(k)$ be Gaussian

$$f(k) = \frac{1}{\sqrt{2\pi}\sigma_k} \int exp\{i(k - k_0)/2\sigma_k{}^2\} dk, \qquad (11)$$

where is the wave vector at the center of the Gaussian profile. Inside the optical media the wave packet pulse is transformed in an evanescent Gaussian pulse, where each component forming the packet is a non-propagating wave with imaginary wave vector [36]. Therefore, our model can be schematized as follows

$$\begin{cases} E_z(x < 0) = \int f(k) exp\{i(\omega t - kx)\} dk - R^{1/2} \int f(k) exp\{-i(\omega t - kx)\} dk \\ D_z(0 \leq x \leq L) = \theta(t) \int f(\chi) exp\{-|\chi|x\} d\chi \\ E_z(x > L) = \theta(t - \tau) T^{1/2} \int f(k) exp\{i[\omega(t - \tau) - k(x - L)]\} dk \end{cases} \qquad (12)$$

where $\theta(t)$ is the Heaviside step function, such that $\theta(t) = 0$ as $t < t_0$ and $\theta(t) = 1$ as $t > t_0$ ($t_0$ is the time at $x = L$), $\chi$ is the imaginary wave vector and $f(\chi)$ is the Gaussian envelop function for the evanescent wave packet, $R$ and $T$ are the reflection and transmitted coefficients respectively. The time $\tau$ is the time needed by the photon to travel from $x = 0$ to $x = L$, but it cannot be confused with the tunnelling time, a meaningful quantity that is not necessary in the theory being formulated. The wave vector of the evanescent mode is given by

$$\chi = \omega\sqrt{\varepsilon(x)}/c, \qquad (13)$$

where $\varepsilon(x)$ is the function representing the permittivity of the optical material forming the waveguide. Since $\chi$ is imaginary then $\varepsilon(x)$ must be negative. Negative values of $\varepsilon$ are found in innovative materials, such as nano-particle materials and metamaterials, which are used in superconducting and signal transmission technology [37-38]. However, a negative $\varepsilon(x)$ violates the Lorentz invariance, and the following equation no longer holds

$$\partial_x^2 E_z - \frac{\varepsilon(x)}{c^2}\partial_t^2 E_z = 0. \qquad (14)$$

Therefore, this situation is not compatible with the theory we are developing, the goal of which is to comply with Lorentz invariance.

## 4. Photon Tunnelling: Recovering the $\varepsilon(x)$ Positive Sign Through Spacetime Complexification

PG solved this problem mapping the spacetime as follows

$$t \to -it \quad and \quad x \to -ix. \qquad (15)$$

This approach is unusual in relativistic quantum mechanics and is widely applied in string theory as it allows for describing spaces of dimensions greater than four. Notably, spacetime complexification can be performed using the Kähler manifold [39]. In this way is proved that any two observers $S$ and $S'$ moving with relative velocity $u < c$, *see* spacetime as a complexified 4-dimensional manifold, where the laws of nature remain unchanged under the action of the elements of the Lorentz group. Notably, Maxwell's equations, in a vacuum, are invariant under the transformations given by Eq. (15). This result has been known for some time and is detailed in the references [40-41]. Maxwell's equations, in media, are also invariant, provided $\varepsilon(-ix) = \varepsilon(x)$. In fact, Eq. (14) is equal to its complexified form, obtained by substituting, in each term, the real variables with the complex variables given by Eq. (15), if $\varepsilon(-ix) = \varepsilon(x)$. As mentioned, each component of the wave function ($\Gamma\psi$) separately satisfies Maxwell's equations. Furthermore, the Harish-Chandra equation, whose linear form recalls that of Dirac, is invariant under the transformations of Eq. (15) [42]. Therefore, our theory continues, in a complexified spacetime, to remain consistent with the goals we have set from the beginning. It is important to note that, using a mapping of Eq. (15), the points on the light cone remain on the light cone. This better clarifies why, under this mapping, the relativistic equations introduced to this point remain valid, being formulated for massless spin-1 bosons.

Let us now suppose that $\varepsilon(x)$ varies slowly along the direction of the x-axis so that it can be developed by the Taylor series about $x = 0$ as follows

$$\varepsilon(x) \approx \varepsilon_0 + \sum_{n=1}^{\infty} a_n x^n, \qquad (16)$$

where $\varepsilon_0$ is the dielectric constant in vacuum, which is set as equal to one. As we will show in the next section, the propagation of the photon inside the barrier is superluminal if $\varepsilon(-ix)$ is positive but less than one. This constraint holds if the infinite sum in Eq. (24) is within the range of $(-1,0)$, and this is systematically verified in a complexified spacetime. In fact, using the transformations of Eq. (15) and the condition for which $\varepsilon(-ix) = \varepsilon(x)$, we can rewrite Eq. (16) as:

$$\varepsilon(-ix) \approx \varepsilon_0 + \sum_{n=1}^{\infty} a_n (-ix)^n \quad \forall n \mid Im[\varepsilon(-ix)] = 0. \qquad (17)$$

From Eq. (17), we see that all $n$ even are acceptable, since they give a negative contribution to the Taylor expansion. In summary, materials with permittivity $\varepsilon(x) > 1$ in ordinary Minkowski spacetime can have a real value less than one in complexified spacetime, as long as the imaginary component is zero. Our theoretical model will be further constructed based on this result.

The possibility of having dielectric constants which are analytic functions has been known since the early 1900s [43]. The Drude-Lorentz theory has been developed to account for the complex index of refractions and dielectric constants

of materials. This model, for a homogeneous and isotropic media, leads to the following formula

$$\varepsilon(\omega) = 1 - \frac{\omega_c^2}{\omega^2 - i\gamma\omega}, \tag{18}$$

where $\gamma$ is the damping coefficient. The Eq. (18) can be rewritten in a more convenient way (by applying the division formula of complex numbers)

$$\varepsilon(\omega) = 1 - \frac{\omega_c^2}{\omega^2 + \gamma^2} - i\gamma\frac{\omega_c^2}{\omega^3 + \gamma^2\omega}. \tag{19}$$

Since our model provides a non-dispersive and non-absorptive media, we can apply the approximation $\gamma \ll 1$, and Eq. (19) becomes:

$$\varepsilon(\omega) \cong 1 - \frac{\omega_c^2}{\omega^2 + \gamma^2}. \tag{20}$$

To ensure that the transmission of the pulse within the waveguide occurs by tunnelling, it is necessary that $\omega$ is lower than the cutoff frequency $\omega_c$. But to have $\varepsilon(\omega)$ less than one but still positive it is likewise necessary that $(\omega^2 + \gamma^2) > \omega_c^2$. Therefore, once the material with $\gamma \ll 1$ has been chosen, the superluminal (Lorentz invariant) tunnelling takes place if the pulse frequency $\omega$ is around the cut off value.

Let us compare Eq. (19) and Eq. (17) whose Taylor development is truncated at the power $n = 2$

$$1 - a_1(-ix) - a_2(-ix)^2 = 1 - \frac{\omega_c^2}{\omega^2 + \gamma^2} - i\gamma\frac{\omega_c^2}{\omega^3 + \gamma^2\omega}. \tag{21}$$

It is clear that this equality holds if the material is homogeneous, i.e. if the functions $a_n(-ix)^n$ are numerical constants. Otherwise, the terms in the second member of Eq. (21) will be functions of the variable $x$ through the spatial dependence of the cut off frequency $\omega_c(x)$ and the damping coefficient $\gamma(x)$. As can be seen, $\varepsilon(-ix) = \varepsilon(x)$ only if $\varepsilon(-ix) \ll 1$. We have therefore given a physical interpretation to the requirement of Eq. (17). In other words, to ensure that the imaginary terms of Taylor's expansion of Eq. (17) are negligible, it is necessary that the optical media has a sufficiently low damping coefficient. This is possible if epsilon-near-zero materials (ENZ) are used, whose dielectric constant is less than one but still positive [44-48].

Using Eq. (20), the explicit form of the modulus of imaginary wave vector $\chi(x)$ for a non-homogeneous dielectric ca be obtained

$$|\chi(x)| = \left(\omega^2 - \frac{\omega^2 \omega_c^2}{\omega^2 + \gamma^2}\right)^{1/2} / c. \quad (22)$$

To obtain a transmitted pulse with not too attenuated intensity, is important that the profile of the transcendent modes in the optical material decrease as slowly as possible. This is possible if the modulus of the wave vector is near to zero, i.e. if

$$\left(\omega^2 - \frac{\omega^2 \omega_c^2}{\omega^2 + \gamma^2}\right) \ll 1. \quad (23)$$

By Eq. (23) is possible to calculate the optimal frequency $\omega$ to ensure superluminal tunnelling once the ENZ material is chosen. The above discussed allows giving form to the idea of PG, laying the operational bases for the construction of an optical device for the experimental verification of the theory. The Drude-Lorentz model declined in the framework of the theory being formulated has allowed identifying the best dielectric medium to be used to achieve the superluminal transport of electromagnetic energy.

**5. Photon Tunnelling: Toward Superluminality**

Now, let us return to where we left off at the end of § 3. The tunnelling mode being investigated is non-stationary. This is the why, in Eq. (12), inside the optical barrier, there is no reflected component of the wave function. The transmitted component of the wave for $x > L$ is always zero, except when $t > \tau$. To get the average transport velocity of electromagnetic energy, the Lorentz invariant formula of Eq. (9) must be used. Since we are considering a one-dimensional problem, the energy density $\psi^t \Gamma \psi / 2$ and the component $T_{01}$ have to be calculated. The energy density $\rho_E (0 \leq x \leq L)$ is obtained by

$$\rho_E = \frac{1}{2} \psi^t \Gamma \psi = \frac{1}{2} \left[\varepsilon(-ix) D_z^2 (0 \leq x \leq L) + B_y^2 (0 \leq x \leq L)\right], \quad (24)$$

where the magnetic field vector inside the optical material is calculated as

$$B_y(0 \leq x \leq L) = \theta(t)\frac{c}{\omega_0}\partial_x D_z(0 \leq x \leq L). \tag{25}$$

The explicit form of function $D_z(0 \leq x \leq L)$ is given by the second of Eq. (12). To calculate the term $T_{01}$ we recall that $mc^3(\psi^t \Gamma \beta_1 \Gamma \psi) = -cD_z B_y = S_x$, where $S_x$ is the x-component of the Poynting vector. Using Eq. (25) and the second of Eq. (12) is obtained

$$S_x = -\frac{c^2}{\omega_0}\left(\int f(\chi)exp\{-|\chi|x\}d\chi\right)\partial_x\left(\int f(\chi)exp\{-|\chi|x\}d\chi\right) \quad 0 \leq x \leq L. \tag{26}$$

Therefore, using Eq. (24) and Eq. (26) we obtain

$$u_x = \frac{S_x(0 \leq x \leq L)}{\rho_E(0 \leq x \leq L)} = \frac{c}{\sqrt{\varepsilon(-ix)}}. \tag{27}$$

Eq. (27) states that, if the material that, in a complexified spacetime, presents the real part of permittivity $\varepsilon(-ix)$ within the range $(0,1)$, and if $Im\varepsilon(-ix) = 0$, then the tunnelling is superluminal. In the previous section we demonstrated that this occurs if ENZ non-absorptive, non-dispersive and non-magnetic materials are used. In particular, using the x-dependent form of Eq. (20) the average velocity becomes

$$u_x = c\frac{\omega^2 + \gamma^2}{(\omega^2 - \omega_c^2) + \gamma^2}, \tag{28}$$

from which we see that has physical meaning as long as $\gamma^2 > (\omega^2 - \omega_c^2)$. Eq. (28) also clarifies why the superluminality condition is obtained by using electromagnetic waves whose frequency is very close the cutoff frequency of the material used.

It must be noted that the invariance of Eq. (27) is ensured by the fact that the matrices $\beta_x$ and $\Gamma$ do not depend on the reference frame; moreover, the components of the transformed wave function separately satisfy Maxwell's equations in complexified spacetime, as mentioned in the previous section.

Experimental demonstrations of superluminal light propagation in ENZ materials has been reported, but not in the framework of quantum tunnelling [48]. The velocity in Eq. (28) depends only on the physical properties of the material and does not provide any information about the geometry of the optical device. This is another peculiarity that distinguishes our proposed theory from non-relativistic theories, where the superluminal behaviour emerges for sufficiently large barriers [49], as the Hartmann effect prescribes [50]. We believe that this result is relevant for designing new optical devices, since thin optical barriers are sufficient to boost the transport of electromagnetic energy. Due to the optical barrier, the signal attenuation remains the main problem; its initial power must be sufficiently high to survive until it reaches the end of the barrier. The maintenance of this energy flux depends on the waveguide, which must be made of low-dispersion material, having a refractive index relative to the frequency of the radiation carrying the energy, very close to one [50]. This material must also be able to maintain the polarisation and coherence of the initial impulse. Therefore, an ideal optoelectronic device for experimentally verifying the possibility of the superluminal transport of electromagnetic energy will be engineered to have a low refractive index waveguide, an ENZ optical barrier and a coherent optical pulse amplifier. In particular, the structure of such devices must be periodic to compensate for the dispersion of the signal in the waveguide and to maintain the intensity of the initial signal.

The formulated theory provides an average velocity and not an instantaneous one. To be reasonably sure that the photonic tunnelling is superluminal is necessary to demonstrate that in each region of the waveguide the instantaneous velocity is superluminal.

## 6. Photon Tunnelling: Theory Validation

To calculate the instantaneous relativistic velocity of the photon inside the waveguide, let us go into the framework of quantum electrodynamics. The method consists in calculating the probability that the photon propagates inside the dielectric medium under the hypothesis that the path is spacelike. If this probability is non-zero whatever the relativistic path is, then the photon velocity (even the instantaneous one) is greater than the speed of light.

Let us denote by $\psi(x^\mu)$ the bosonic field representing the photon, where $x^\mu = (t, x, y, z)$. Inside the optical medium the field $\psi(x^\mu)$ is represented by $D_z$, which now assumes the form of quantised field operator (we are always considering the one-dimensional problem addressed in the previous sections). Let us denote by $x^\mu = (t, x, 0, 0)$ and by $x'^\mu = (t', x', 0, 0)$ two spacetime points inside the waveguide. The photon transition probability amplitude associated to the path $\Delta s = (x^\mu - x'^\mu)$ is given by

$$G(x^\mu - x'^\mu) = \langle 0|D_z(x^\mu)D_z(x'^\mu)|0\rangle, \tag{29}$$

where $|0\rangle$ here is the vacuum state. The square of function $G(x^\mu - x'^\mu)$, i.e. $|G(x^\mu - x'^\mu)|^2$ is the probability that the photon propagates from $x^\mu$ to $x'^\mu$. We suppose that this this is a spacelike path, i.e. $(\Delta s)^2 < 0$. Considering that the field operator $D_z$ inside the waveguide is evanescent and applying the theory of quantum electrodynamics we obtain

$$G(x^\mu - x'^\mu) = \left(\frac{\partial^2}{\partial z^2} - \frac{\partial^2}{\partial t^2}\right)\frac{\varphi}{4\pi}\int \frac{e^{[-i\omega(t-t')+i\chi(x-x')]}}{\omega}d\chi, \tag{30}$$

where $\varphi$ is a factor phase that, for our purpose, is not necessary to know explicitly. Since the integral kernel does not depend on $z$, the derivative $\partial_z^2$ is zero and Eq. (30) becomes

$$G(x^\mu - x'^\mu) = -\frac{\partial^2}{\partial t^2}\frac{\varphi}{4\pi}\int \frac{e^{[-i\omega(t-t')+i\chi(x-x')]}}{\omega}d\chi, \tag{31}$$

The wave vector $\chi$ is a function of $\omega$ and $x$, therefore using Eq. (13) we get

$$d\chi = \frac{\sqrt{\varepsilon(x)}}{c}d\omega. \tag{32}$$

For simplicity we can assume that $\varepsilon(x)$ is constant within the considered path. Substituting Eq. (13) and Eq. (32) in Eq. (31) is obtained

$$G(x^\mu - x'^\mu) = -\frac{\partial^2}{\partial t^2}\frac{\varphi}{4\pi}\int \sqrt{\varepsilon(x)}\frac{e^{\left[-i\omega(t-t')+\left|\sqrt{\varepsilon(x)}\right|\omega(x-x')/c\right]}}{\omega c}d\omega. \tag{33}$$

Since the tunnelling occurs for $\omega < \omega_c$, the extremes of integration in Eq. (33) become $\omega_c$, the upper one, and $\omega > 0$ the lower one. The analytic function $\varepsilon(x)$

depends on the frequency $\omega$, as detailed in Eq. (20); therefore Eq. (33) can be rewritten as

$$G(x^\mu - x'^\mu) = -\frac{\partial^2}{\partial t^2}\frac{\varphi}{4\pi}\int i\left(1 - \frac{\omega_c^2}{\omega^2 + \gamma^2}\right)\frac{e^{\left[-i\omega(t-t\prime)+1-\frac{\omega_c^2}{\omega^2+\gamma^2}\omega(x-x\prime)/c\right]}}{\omega c}d\omega. \qquad (34)$$

The integral of Eq. (34) is quite complicated but we do not have to solve it exactly: we have only to prove that is not equal to zero. Since $0 < \omega < \omega_c$, the integral kernel does not ever vanish. Furthermore, its temporal dependence is given by an exponential function which, even if repeatedly derivate with respect to $t$, will never become zero. This is enough to state that $G(x^\mu - x'^\mu)$ is always different from zero, also for infinitely small spacelike paths $\Delta s$. Therefore, we have proved (using a Lorentz invariant theory) that the velocity of the photon inside the waveguide is always superluminal, also for infinitesimal path (which implies that also the instantaneous velocity is always greater than the speed of light).

A similar result has been obtained also by Nimtz, considering the negative energy of an evanescent field inside a waveguide [51].

## 7. Tunnelling Photon: How Preserving the Causality Principle

We have developed a theory that corroborates the possibility of transporting electromagnetic energy at superluminal velocity in suitable optical devices. But this inevitably leads us into the insidious framework of the causality paradox, a problem physicists of tachyons know have to address constantly. Considering our one-dimensional model, the problem is: an observer in a subluminal reference frame see an electromagnetic signal emitted at $(t, x, 0,0)$ and absorbed at $(t', x', 0,0)$, where $[(x^2 - x'^2) - c^2(t^2 - t'^2)] > 0$. A second observer in a superluminal reference frame see the electromagnetic see the particle absorbed at $(t', x', 0,0)$ before it is emitted at $(t, x, 0,0)$. Invoking the Feinberg reinterpretation principle [52], things return to be consistent with the physical reality: *tachyon sent back in time can always be reinterpreted as a tachyon traveling forward in time, because observers cannot distinguish between the emission and absorption of tachyons*. In other words, the second observer sees a signal emitted at $(t',x',0,0)$ and absorbed at $(t,x,0,0)$. But the signal seen by the second observer is necessary different from the one seen by the subluminal observer. More precisely, the second observer sees the antiparticle of the particle seen by the first observer. This is deepest discussed in the fundamental paper by Recami [53]. In our model the tunnelling photon is investigated, and the photon is its own antiparticle. Therefore, in the framework of reinterpretation principle, the two observers see exactly the same phenomenon, which solve the problem of causality of our theory.

**8. Concluding discussion**

Superluminal behaviours of matter or light have always attracted the interest of physicists, mostly arousing scepticism and creating controversy. Theoretical physics has long demonstrated that such phenomena can be introduced within the framework of the theory of relativity, without undermining its foundations. Using the classical theory of electromagnetism, for example, has revealed that an electromagnetic pulse can be transmitted at superluminal speed in a linear dielectric medium [54-55]. These theories are Lorentz invariant, and they preserve the principle of causality as long as the medium exchanges energy asymmetrically with the leading and trailing portions of the pulse. However, for phenomena in which the electromagnetic pulse is transmitted through tunnelling, an invariant Lorentz theory, in which the principle of causality holds, is still lacking, and this cannot be formulated in the framework of classical electrodynamics. To this purpose, we formulated a Lorentz invariant quantum model that is faithful to the theory of relativity and does not require any other speculative assumptions. This has been done starting from a work by PG [31] who, although little considered in the framework of photonics, brilliantly solves the problem of the relativistic invariance of superluminal tunnelling.

The formalism used allows investigation of the tunnelling of a finite width photon wave packet incident normally on a one-dimensional photonic barrier. To the best of our knowledge, this approach is innovative because, in optics, the total internal reflection occurs only for non-zero critical angles of incidence, while this constraint breaks down in our calculations. The study has clearly shown that Einstein causal superluminal propagation can occur only if the tunnelling medium is non-absorptive, non-dispersive, non-magnetic and (but not streakily necessary) inhomogeneous, and if $Im[\varepsilon(-ix)] = 0$. This is a consequence of the spacetime complexified mapping introduced to solve the problem of negative permittivity, which is not compatible with Eq. (14). Therefore, to obtain a superluminal signal, it is necessary to use a dielectric medium with $\varepsilon(x) > 1$ in ordinary Minkowski spacetime. However, once mapped in complexified spacetime, this leads to $\varepsilon(-ix) < 1$ with $Im[\varepsilon(-ix)] = 0$, ensuring the relativistic invariance of Eq. (21). These materials are defined as *ultrarefractive*, and, near their characteristic cut off frequency, the permittivity becomes less than one [56-59]. Such materials have been used in experiments discussed in references [8, 10], where superluminal tunnelling occurs at near normal incidence through the band gaps inherent in periodic dielectric structures. These band gaps arise from the Bragg reflections, which lead to evanescent decay of the wave function as soon as the frequency is within the classically forbidden band gap in the first Brillouin zone. The periodic structure of these ultrarefractive materials makes them non-dispersive so that the wave function, after tunnelling, remains undistorted, even if it is attenuated in amplitude. The repetition of these experiments with more sophisticated devices, designed with periodic sequences of ENZ optical media and coherent optical pulse amplifiers on low refractive index waveguides, will be the Galilean test that could confirm or refute the possibility of electromagnetic energy transport at velocities greater than $c$.